\documentclass[5p,times]{elsarticle}

\usepackage{hyperref} 
\usepackage{amsmath}
\usepackage{amsfonts}
\usepackage{amssymb}
\usepackage[dvipsnames]{xcolor}
\usepackage[xcolor]{mdframed}

\newcommand{\brkival}[2]{[#1\,\colon\,#2]}
\usepackage[]{algorithm2e}
\newcommand{\vect}{\vec}
\journal{Arxiv}
\newcommand{\rdf}{g}
\newcommand{\rcdf}{G}
\newcommand{\graph}{H}
\newcommand{\pedsq}{ped/m$^2$}

\usepackage{subcaption}

\begin{document}

\begin{frontmatter}
\title{Monitoring physical distancing for crowd management:\\
real-time trajectory and group analysis}
\author[AP,ProRail]{Caspar A.S. Pouw}

\author[AP,CNR-IAC]{Federico Toschi}
\author[ProRail]{Frank van Schadewijk}

\author[AP]{Alessandro Corbetta \corref{mycorrespondingauthor}}
\cortext[mycorrespondingauthor]{Corresponding author}
\ead{a.corbetta@tue.nl}

\address[AP]{Department of Applied Physics, Eindhoven University of Technology 5600 MB Eindhoven, The Netherlands}

\address[ProRail]{ProRail Stations, 3511 EP Utrecht, The Netherlands}
\address[CNR-IAC]{CNR-IAC I-00185, Rome, Italy}

\begin{abstract}
  Physical distancing, as a measure to contain the spreading of Covid-19,
is defining a ``new normal''. Unless belonging to a family,
pedestrians in shared spaces are asked to observe a minimal
(country-dependent) pairwise distance. Coherently, managers of
public spaces may be tasked with the enforcement or monitoring of this constraint.
As privacy-respectful real-time tracking of pedestrian dynamics in
public spaces is a growing reality, it is natural to leverage on
these tools to analyze the adherence to physical distancing and
compare the effectiveness of crowd management measurements.  Typical
questions are: ``in which conditions non-family members
infringed social distancing?'', ``Are there repeated offenders?'', and ``How are new crowd management measures performing?''.  Notably, dealing with large
crowds, e.g.\ in train stations, gets rapidly computationally
challenging.  
 
In this work we have a two-fold aim: first, we propose an efficient and scalable analysis
framework to process, offline or in real-time, pedestrian tracking
data via a sparse graph. The framework tackles efficiently all
the questions mentioned above, representing pedestrian-pedestrian interactions via vector-weighted graph connections.
On this basis, we can disentangle distance offenders and family members in a privacy-compliant way.
Second, we present a thorough analysis of mutual distances and exposure-times in a Dutch train platform, 
comparing pre-Covid and current data via physics observables as Radial Distribution Functions.
The versatility and simplicity of this approach, developed to analyze crowd management
measures in public transport facilities, enable to tackle issues beyond physical distancing, for instance the
privacy-respectful detection of groups and the analysis of their motion patterns.
\end{abstract}


\end{frontmatter}


\section{Introduction}
Crowd management is a challenging scientific topic directly impacting on the functioning of trafficked urban infrastructures such as, e.g., train or metro stations. Even more so, in time of Covid-19 pandemic, after an initial lock-down period, communities are still wondering how to resume a ``new normal'' life, while the virus is still circulating among the population. One of the key control measures has been to maintain a minimal physical distance (often also called ``social distance'') between any two individuals not belonging to the same family~\cite{who-pd}. This distance is country-specific and it ranges from $1\,$m (e.g.\ China and France), as recommended by WHO, up to $2\,$m (e.g.\ UK and Canada), being $1.5\,$m in the Netherlands and in some countries it is even adjusted over time. As there is a rather widespread suspicion that we may have to live with such requirements of physical distancing for months to come,  it is therefore natural that this is becoming a design requirement for public infrastructures (e.g.~\cite{Tian2020AnChina,Rader2020CrowdingTransmission,anderson2020will,chinazzi2020effect}). 

There are however several challenges associated to the automated monitoring of physical distancing in crowds. First, in order to respect individual privacy one needs to employ sensors and techniques that ensure privacy by design while, at the same time, providing accurate space-time information on individual positions with sub-meter accuracy.

Secondarily, one needs to develop algorithms that, while preserving privacy, are capable to autonomously discern, with a good degree of accuracy, families and family members from strangers. This identification should be performed in real-time, raising a number of non-trivial technical challenges.

Additionally, in recent months a number of countries have developed contact tracing apps that allow to receive an alert when somebody has been in ``close'' contact with somebody that, later on, will turn out to be positive to the Covid-19~\cite{apple-google}. Countries are developing apps based on different alert thresholds, typically a combination of having been closer than a given distance, for a time longer than an established reference. These thresholds, again, are country specific. In Italy and Germany the national apps alert for contacts longer than 15 minutes at a distance below 2 meters.  The proper balance of these two aspects, distance and time, is key to avoid too many false positives or false negatives due to e.g.\ to low or high risk exposures as well as to the inaccuracy of distance estimation via the intensity of Bluetooth signals. It is therefore extremely interesting to be able to analyze, in a number of key urban settings, the combination of contact times for  given distances between two persons. This knowledge may provide key information for the calibration of contact tracing apps in different context.

\begin{figure*}[ht]
    \centering
    \includegraphics[width=\linewidth]{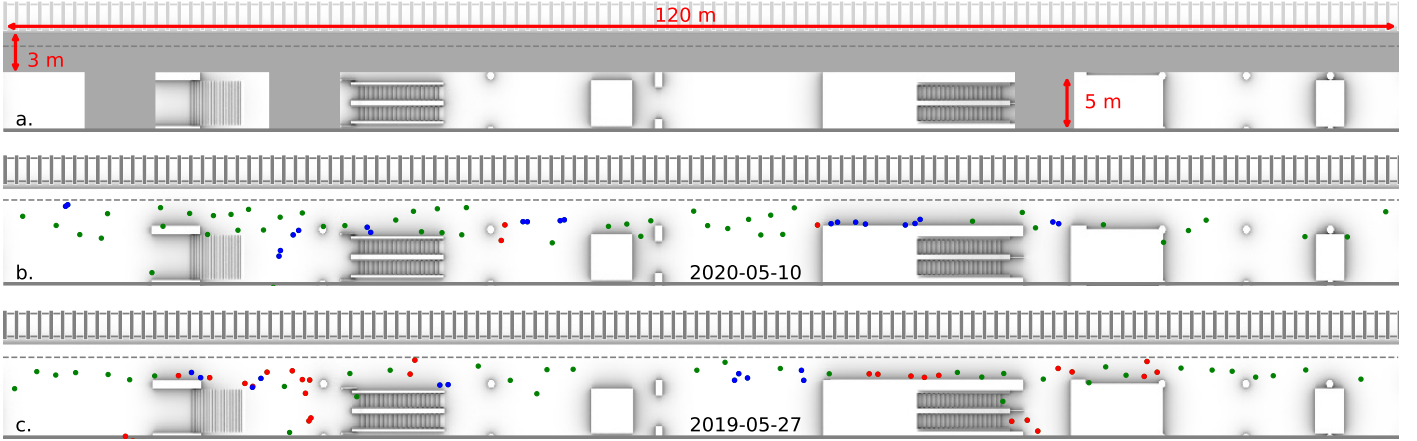}
    \caption{ (a) Floorplan of platform 3 at Utrecht Central Station (NL). The area monitored by the sensors is highlighted in grey. (b) Sample of 75 passengers waiting for a train to arrive on the 10$^{th}$ of May 2020, during the Covid-19 pandemic. Pedestrians which respect the $1.5\,$m physical distance regulations are colored in green, people who are part of a family-group are colored in blue, distance offenders are colored in red. This classification is performed via the method proposed in Section~\ref{sec:method}. In this situation only 3 out of the 75 people violate the physical distancing rules. (c) Same number of people distributed over the platform on the 27$^{th}$ of May 2019, one year prior to the Covid-19 outbreak, here about one-third of the people stand closer than $1.5\,$m to someone else.}
    \label{fig: dist_on_pf}
\end{figure*}

In this paper we employ data from commercial pedestrian tracking sensors placed overhead at Platform 3 in Utrecht central train station (The Netherlands), in order to develop an efficient algorithm, capable of running in real-time, and able to distinguish infringements of the physical distancing rule from the behaviour of family members, that are allowed not to respect such a rule. We introduce the concept of ``Corona event'', to indicate events when two people, not belonging to the same family, get closer than a threshold distance $D$.

We focus on contact times and mutual distances considering statistical observables as the radial distribution functions (RDFs), which can conveniently be employed to quantify average exposure times. This enables a two-fold task: automatizing the definition of families and groups  (from now on named family-groups) and characterizing the statistical distribution of violations, which we compare with analogous pre-Covid measurements. Based on the space-time dynamics of groups, we try to identify family members as those individuals that consistently stay closer than a given threshold distance for sufficiently long time. This, in turn, allows us to define physical distance violators as those individuals that only occasionally (i.e.\ inconsistently) yield Corona events infringing the minimal distance rule.

This paper is structured as follows: in Section~\ref{sec:related-work}, we survey the pedestrian dynamics literature and computer science methods in connection with group-dynamics and mutual distances. We outline both fundamental outstanding questions and existent analysis methods. In Section~\ref{ref:utrecht-intro}, we describe the location and measurement setup at Utrecht Central train station used to acquire the analyzed pedestrian data. In Section~\ref{sec:RDF} we review the concept and basic properties of Radial Distribution Functions, extensively leveraged on by our method. In Section~\ref{sec:method} we present our method and possible variations, that we employ to analyze mutual distancing data in Section~\ref{sect:results}. A final discussion in Section~\ref{sect:discussion} closes the paper.

\begin{figure*}
    \centering
    \includegraphics[width = 0.5\linewidth]{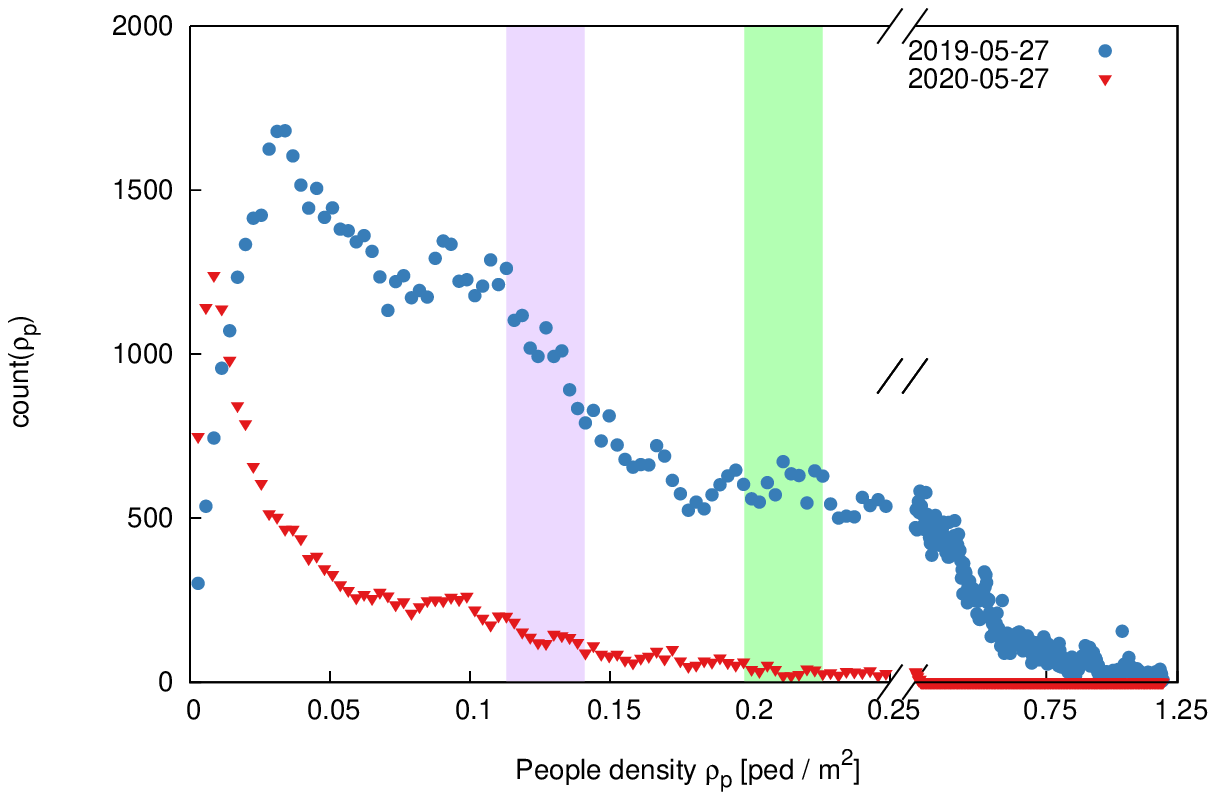}
    \caption{Histogram of observed crowd density levels comparing a day before the Covid-19 outbreak (27$^{th}$ of May 2019, blue dots) and for a day during the Covid-19 pandemic (27$^{th}$ of May 2020, red triangles). Prior to the Covid-19 outbreak, densities in excess of $1\,$\pedsq\ occurred daily. One year later, during the Covid-19 pandemic, the maximum crowd density recorded is only about $0.3\,$\pedsq. We compare measurements acquired at similar density levels, i.e.\ where the average available space per person is comparable. We focus on two density levels: 40-50 passengers (purple band, cf.\ Figures~\ref{fig: dist_hist}(a,c) and~\ref{fig: dist_hist_weekend}(a)) and 70-80 passengers (green band, cf.\ Figures~\ref{fig: dist_on_pf}, \ref{fig: dist_hist}(b,d) and~\ref{fig: dist_hist_weekend}(b)).
    }
    \label{fig: dist_hist_tyical_day}
\end{figure*}

\begin{figure*}
    \centering
    \includegraphics[width = \linewidth]{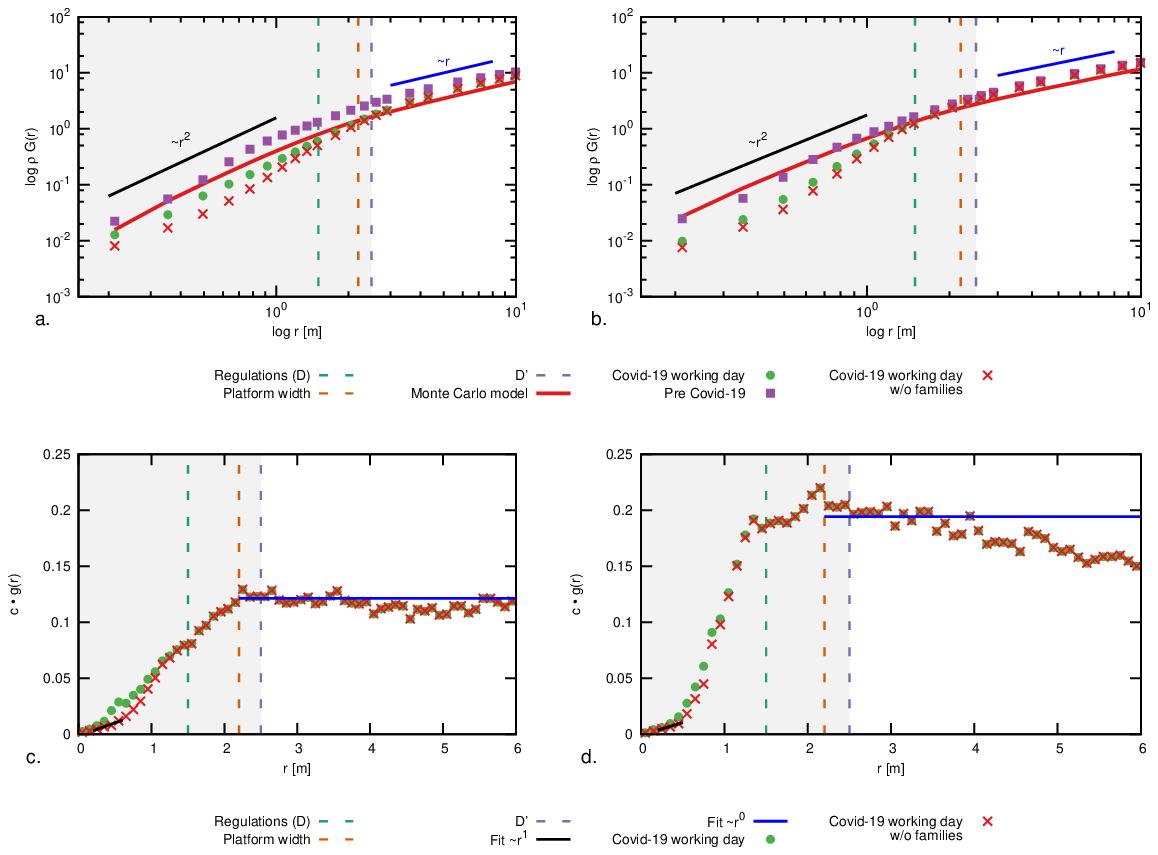}
    \caption{(a, b) Radial cumulative distribution functions (RCDF), $g(r)$, and (c, d) radial distribution functions  (RDF), $G(r)$, for density levels on a typical working day. On the left (a, c) for density level 1, with 40-50 pedestrians on the platform, (green domain in Figure~\ref{fig: dist_hist_tyical_day}) and on the right (b, d) for density level 2, with 70-80 pedestrians on the platform (purple domain in Figure~\ref{fig: dist_hist_tyical_day}). 
    Vertical dashed lines at $1.5\,$m, $2.2\,$m and $2.5\,$m indicate, respectively, the Dutch social distancing regulations ($r<D$), the usable width of the platform (without danger zone) and the critical threshold $D'$.
    The solid black line at small $r$ values highlights the ${\sim} r^2$ growth ratio up to $2.2$ m and a blue line for the ${\sim} r^1$ trend at larger mutual distances. In (b,d) the normalization constant $c$ is chosen such that $\int_0^\infty c g(z)\,dz = N$, where $N$ is the number of people on the platform. Similar plots for a weekend day are reported in Figure~\ref{fig: dist_hist_weekend}. We compare the pre-Covid situation with the present, and with a Monte Carlo model of a random distribution of passengers across a region identical to the platform. We report the RDF and RCDF of the current situation including and excluding family-groups contributions, as made possible by the method introduced in Section~\ref{sec:method}.
    }
    \label{fig: dist_hist}
\end{figure*}

\section{Related works: (social) distance in pedestrian dynamics}\label{sec:related-work}

The analysis of pairwise distances and the automated identification of family-groups triggered by the Covid-19 pandemics connect with outstanding technological and fundamental issues in the broader field of \textit{crowd dynamics}. Crowd dynamics is a multidisciplinary research area aiming at understanding and modeling the motion of pedestrians in crowds (see, e.g.,~\cite{cristiani2014BOOK,helbing2001traffic,adrian2019glossary}, for introductory references). Outstanding questions specifically connected to mutual distances and groups are, e.g.: ``what is the impact of the group on the individual dynamics observables such as position and velocity?'' ``How do people in social groups interact?'', ``How does information propagates throughout groups?'' (see e.g.~\cite{moussaid2010walking,PhysRevE.89.012811,gorrini2016age,doi:10.1080/21582041.2011.619867,zanlungo2020social} and e.g.~\cite{templeton2015mindless} for a group-psychology review). 
Although these questions are longstanding, and have been investigated via models or in laboratory settings extensively, first quantitative studies in pedestrian dynamics driven by real-life big experimental datasets are relatively recent  (see, e.g.,~\cite{Brscic201477,Corbetta2014HighDynamics,corbetta2016continuous,Corbetta2018Physics-basedPedestrians}). Large-volumes of experimental data, in the order of hundred of thousands real-life trajectories, are indeed essential in order to analyze quantitatively and systematically the physics of pedestrian motion, disentangling the high variations in individual behaviors from average patterns, and characterizing typical fluctuations and universal features~\cite{Corbetta2018Physics-basedPedestrians,corbetta2016fluctuations}. This relative delay in performing high-statistics based analyses of pedestrian motion (especially in comparison with other ``active matter'' physical systems~\cite{RevModPhys.85.1143}), is most likely due to the complex technical challenge of achieving accurate, privacy-preserving, individual tracking in real-life conditions (see, e.g., \cite{corbetta2016fluctuations,brscic2013person,seer2014kinects}, or~\cite{KronemanCorbettaPed18} for approaches targeting even higher resolution). Market solutions, as the one considered in this paper, are also becoming accessible, offering various trade-offs between accuracy and costs (see, e.g.,~\cite{Heuvel2019validation}). 

On top of automated tracking, higher-level automated understanding of individual behaviors  -- a concept also known in computer science as trajectory pattern mining~\cite{giannotti2007trajectory,zheng2015trajectory} -- remains also outstanding in many aspects. The automated identification of pedestrian groups, or pedestrian ``group mining'', is a notable example in this context. On one side, in current pedestrian dynamics research, the definition and classification of groups and social structures in experimental data has been manual, i.e.\ based on labor-intensive visual inspection (e.g.~\cite{zanlungo2014pedestrian}). While this ensures high-quality validated measurements, it limits the possibility to establish vast statistical datasets towards data-driven characterizations of averages and fluctuations in the dynamics. On the other side, automatic strategies to identify groups have been proposed by the data mining community. These approaches primarily hinge on analyzing (instantaneous) spatial clusters of pedestrians and the consistency with which these adhere over time to some group semantics (flocking, convoying, aggregation/ desegregation, see~\cite{zheng2015trajectory,benkert2008reporting,vieira2009line,kalnis2005discovering} and references within). 

Here we pursue distance analyses and family-group identification via discretized mutual pairwise distance distributions -- represented in physics terms via Radial Distribution Functions among relevant pedestrian pairs (the concept of RDF is further reviewed in Section~\ref{sec:RDF}). We accumulate information on a ``social'' interaction graph with vector edge weights. This data structure holds all the relevant contact times and distance statistics; besides, family-groups emerge as incremental features queryable  by a space-time distance semantics.
Graphs are classic tools in discrete mathematics to represent networks of interactions, or connections between entities (e.g.~\cite{bondy1976graph}). %
Formally speaking, a graph $H$ is a set of nodes, $H = \{p_i\}$, endowed with edges, say $e = (p_i,p_j)$, connecting node pairs. 
Providing  a weight function, $w(e)$, defined on the edges, makes the graph ``weighted''. In our case, nodes are in 1:1 correspondence with observed pedestrians, whereas edges underlie distance-based interactions, that are characterized by a weight function with values in a real vector space of pre-fixed dimension. 
Graphs have been often used for data-driven studies on social behavior
both of humans, e.g.\ to analyze social
networks~\cite{pitas2016graph}, GPS-data~\cite{guo2010graph}, but also
of social animals
(e.g.~\cite{silk2017understanding,ohtsuki2006simple}).
In~\cite{Corbetta2018Physics-basedPedestrians,Corbetta2017FrameLanding},
graphs have also been used to address big-data analyses and
representation of pedestrian dynamics aiming at efficient data
searches. 

\begin{figure*}[t]
  \centering
    a.\includegraphics[width = 0.48\linewidth]{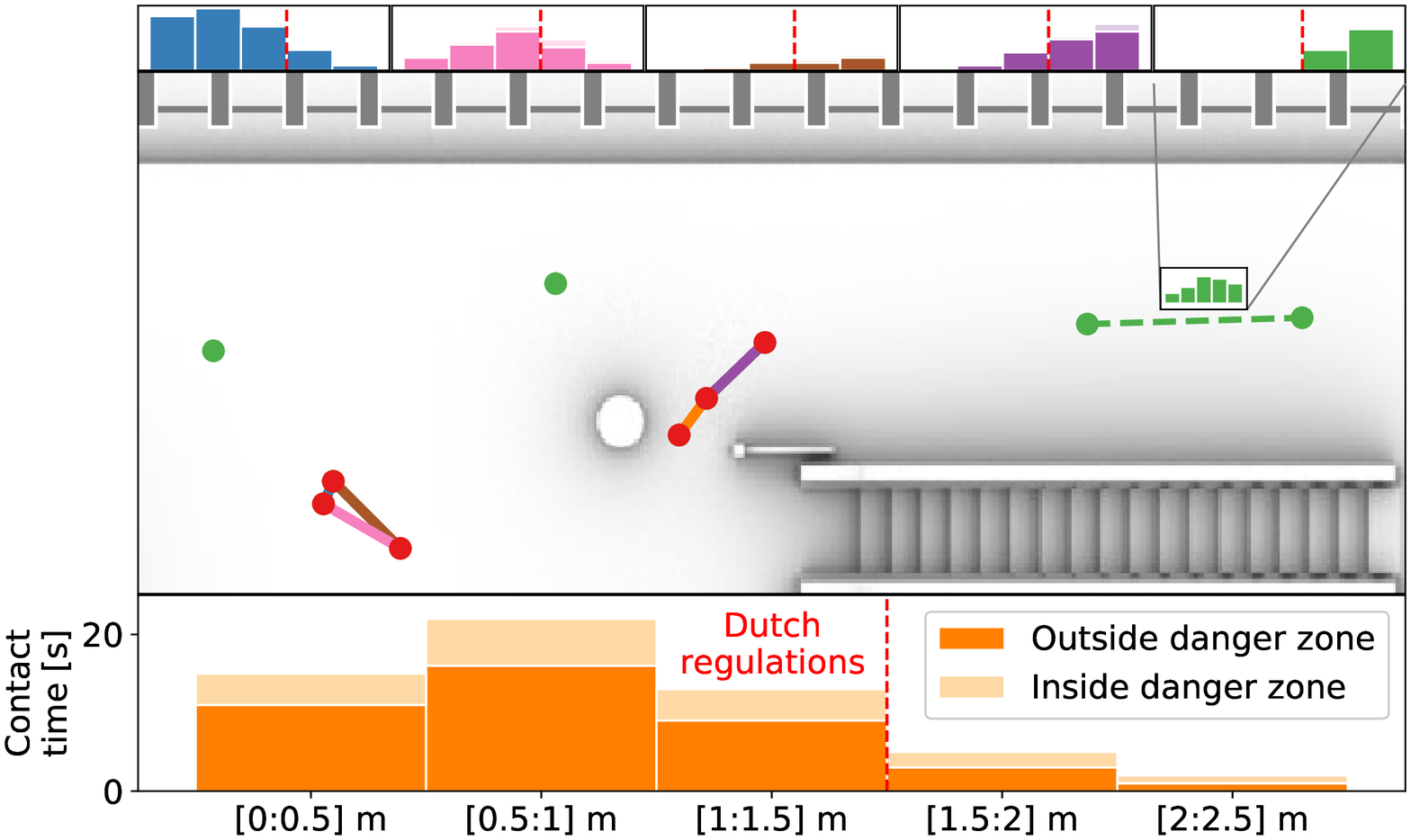}
    b.\includegraphics[width = 0.48\linewidth]{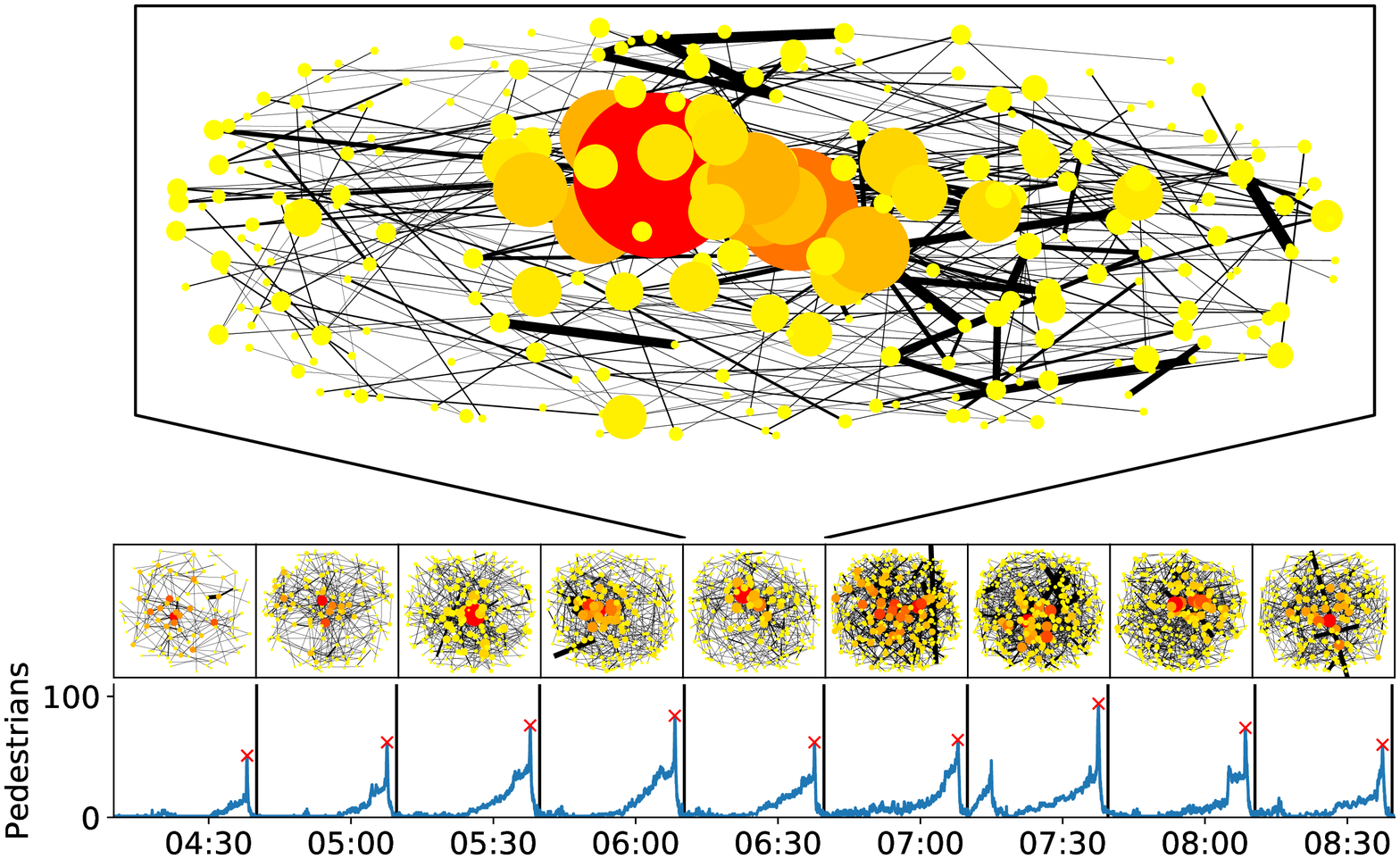}
    \caption{(a) Conceptual sketch representing the accumulation of information on the graph $\graph$. Whenever two pedestrians, say $p_1,p_2$ stand at a distance $d$ smaller than $D'$, this gets recorded in the histogram weight of the edge between nodes $p_1$ and $p_2$ as an additive contribution to the bin approximating  $d$. In the sketch we report a section of the platform: edge appear between nodes according to the distance; the histogram weights are reported atop and beneath the sketch with the same color coding of the edges and scaled with the sampling time (thus they translate to the contact time conditioned by the distance). Nodes are reported in red if they have performed at least one Corona event (thus they have an edge with non-zero contributions at distances below $D'$), else they are in green.
    (b) Examples of graphs acquired in windows of about ten minutes around each train arrival (determining the peaks in the counts at the bottom). We report a magnified version of one among these graphs. Nodes are colored by the node degree, i.e.\ by the number of first neighbors, ranging from yellow to red. Edge thickness scaled by the contact time, $T_e^d$, Eq.~\eqref{eq:pairwise-contact-time}. The higher the degree of a node, the larger the number of distance offenses performed by the associated pedestrian.
    }\label{fig: platform_network_graph}
\end{figure*}

\begin{figure*}[t]
\centering
     a. \includegraphics[width = 0.44\linewidth]{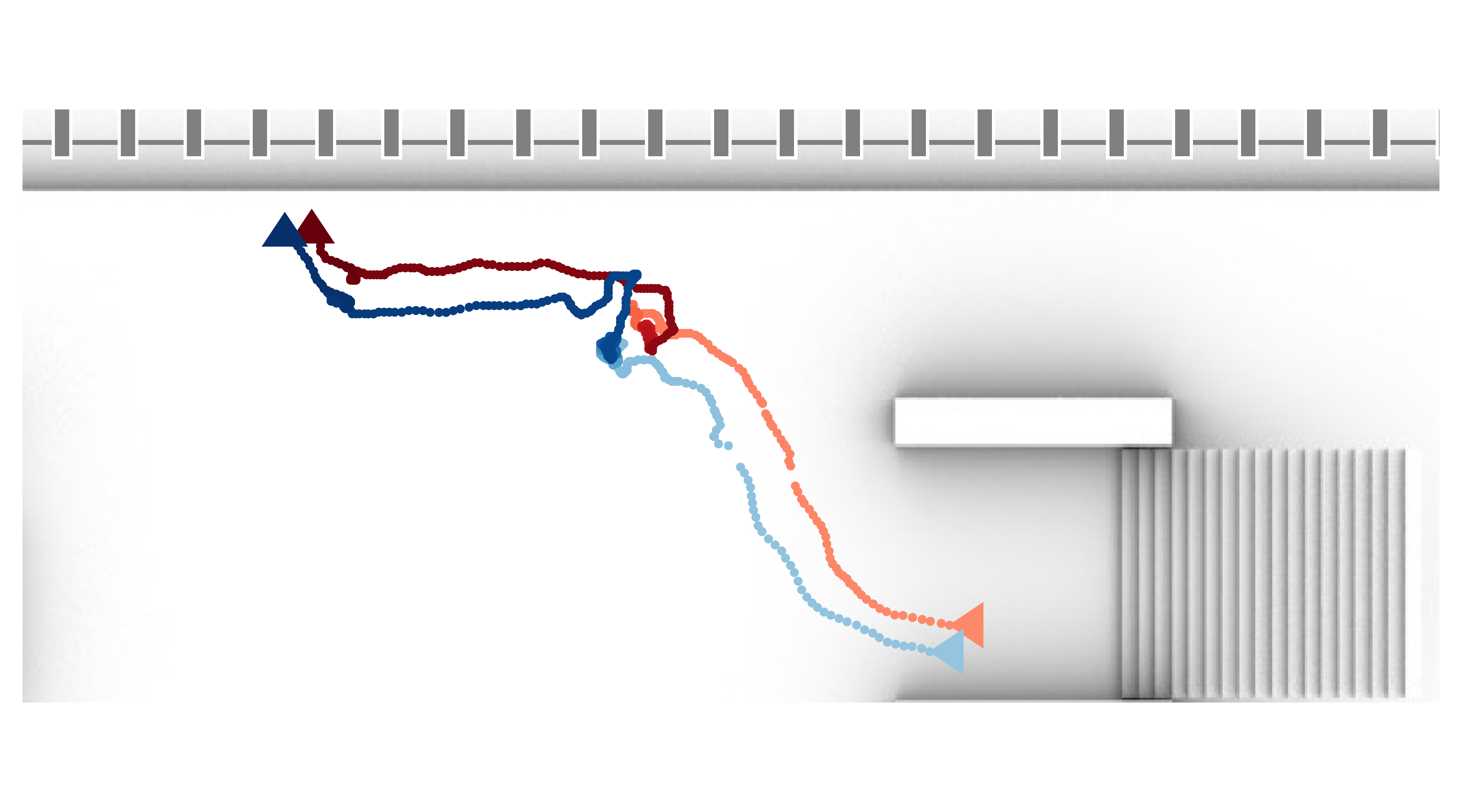}
     b. \includegraphics[width = 0.5\linewidth]{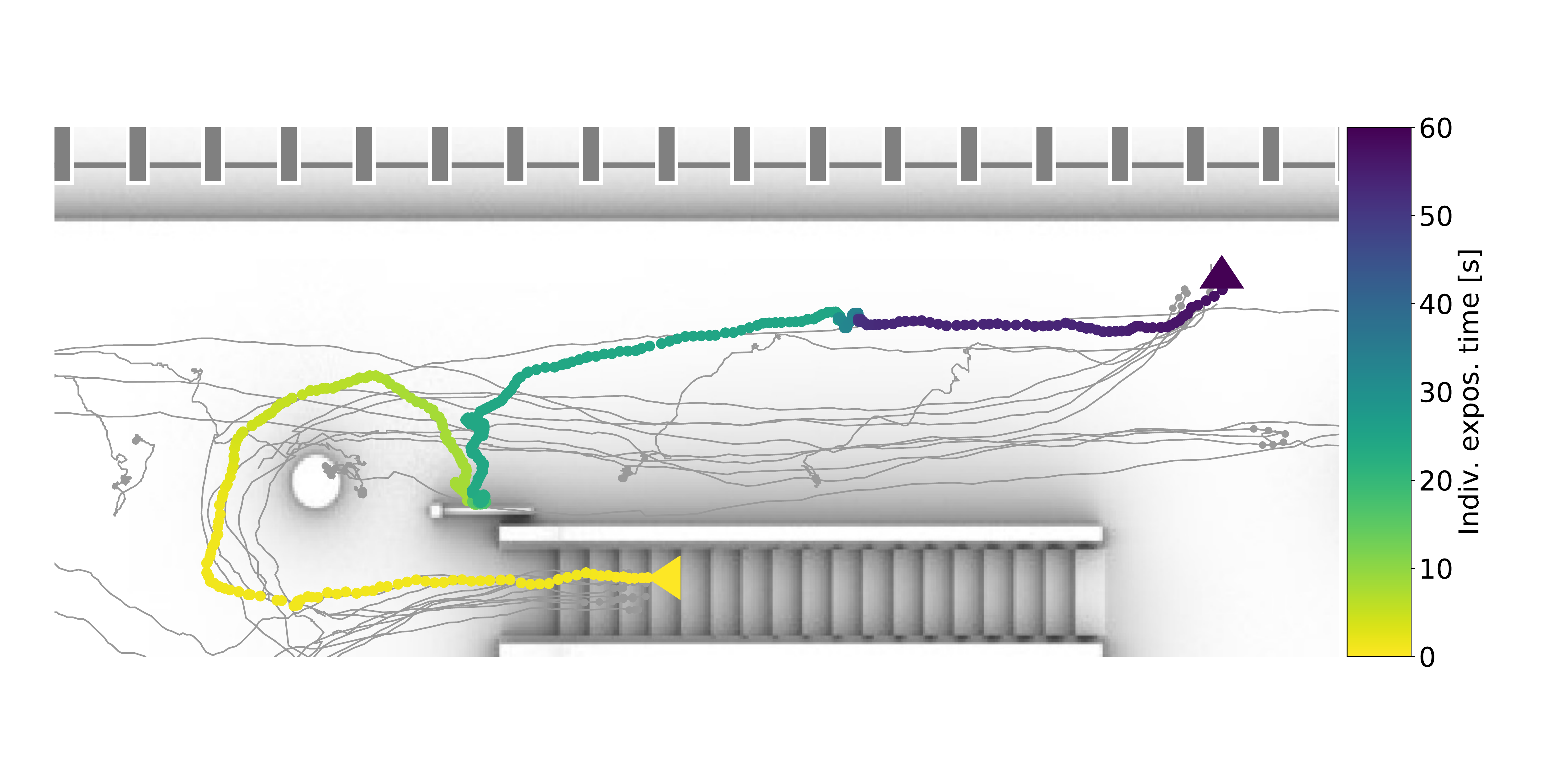}
     
     \caption{(a) Detected clique consisting of two nodes representing two people traveling together. Both entering the platform through the stairs, waiting together for the next train to arrive and finally boarding the train through the same door. The hue of the trajectories is proportional to the time spent on the platform. Lighter hue when the people enter and a darker hue when they leave. Jump in hue indicating the place where the travelers were waiting. (b) Detected node with degree higher than 10, i.e.\ a repeated offender who violates physical distancing with more than 10 other people. The trajectory of the repeated offender is reported in shades scaled to the exposure time, while the trajectories of other people that were met violating physical distancing are in gray. The considered offender entered the platform via the escalators and waited underneath the escalators for their train.}
    \label{fig: duo_offender}
\end{figure*}

\begin{figure*}[t]
    \centering
    \includegraphics[width = \linewidth]{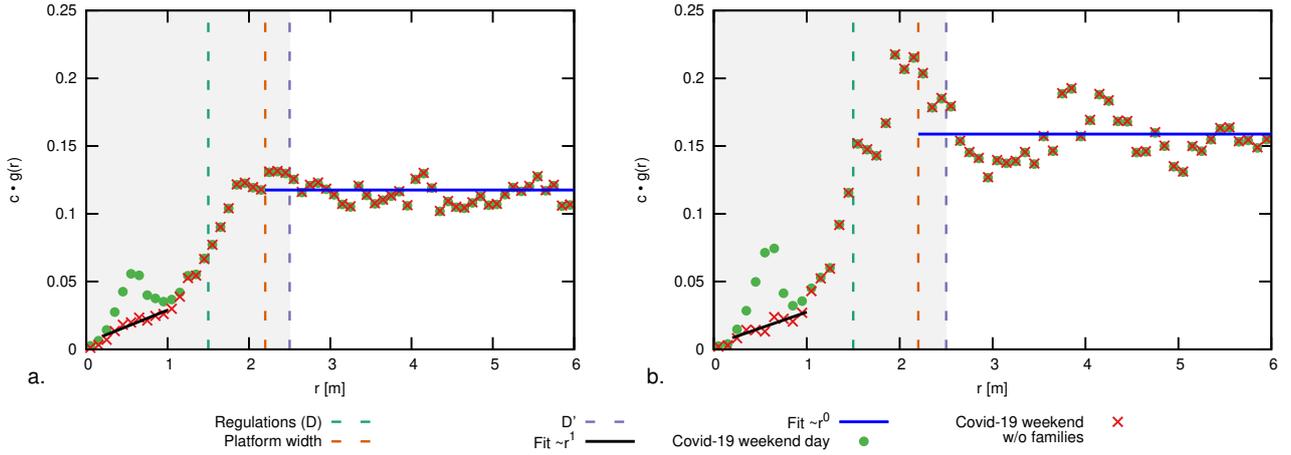}
    \caption{
    Radial distribution functions (RDF), $g(r)$, for a typical weekend day in case of (a) 40-50 pedestrians on the platform, (green domain in Figure~\ref{fig: dist_hist_tyical_day}) and (b) 70-80 pedestrians on the platform (purple domain in Figure~\ref{fig: dist_hist_tyical_day}). The same conventions of Figure~\ref{fig: dist_hist_tyical_day} hold. The presence of family-groups determine a peak in the RDFs around $r\approx 0.5\,$m, which is much more pronounced than in the working day case. Discounting these contributions via the graph analysis notably restores a ${\sim}r^1$ growth rate at small $r$ values.  
    }
    \label{fig: dist_hist_weekend}
\end{figure*}

\begin{figure*}[t]
    \centering
    a. \includegraphics[width=0.43\linewidth]{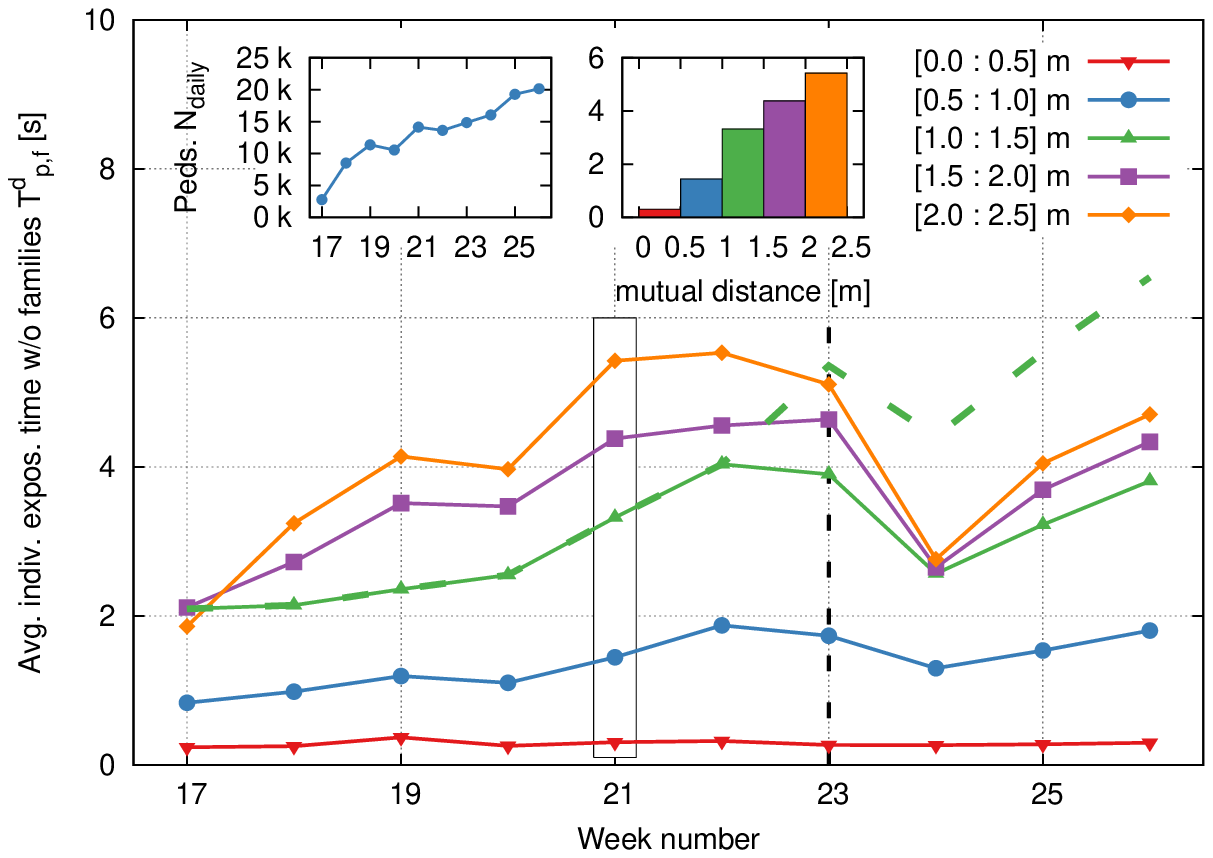} %
    b. \includegraphics[width=0.43\linewidth]{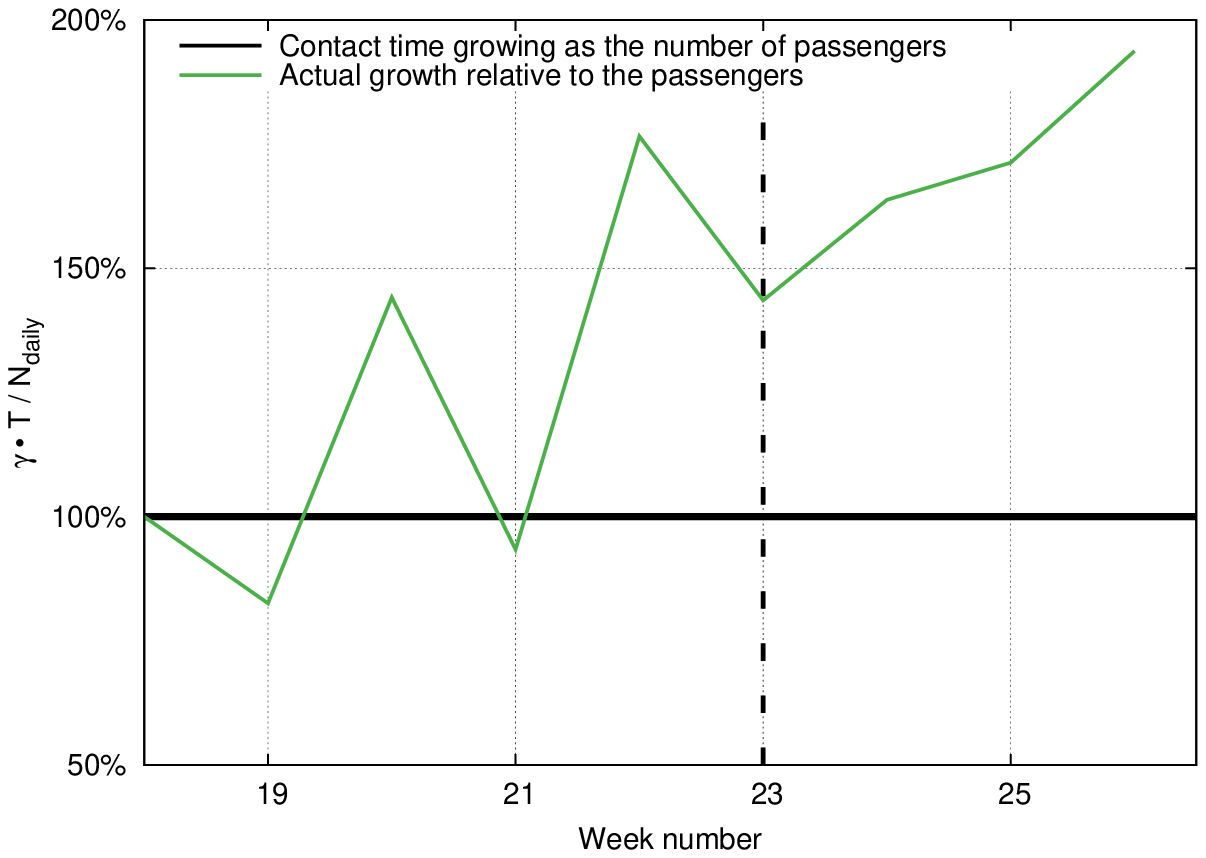}    %
    \caption{(a) Average individual exposure time without family contributions (Eq.~\eqref{eq:total-no-family-contact-time};  weeks 17-26, working days only. Corona lockdown measures in The Netherlands started around week 13). Each line reports average data from bins $\vect{w}_0(e), \vect{w}_1(e)$, etc. The inset on the left shows the average daily passenger count, $N_{daily}$. The inset on the right reports the same individual exposure time data for week 21 in histogram form.
    A change in the train schedule on the 2$^{nd}$ of June (week 23, indicated with a vertical black dashed line) increased the train frequency by a factor $\gamma \approx 1.7$ (Eq.~\ref{eq:correction-fact}). This improved the distribution of pedestrians over the day thereby temporarily decreasing the individual exposure time. 
    To make the data comparable over time and compensate for the train increment, we multiply the exposure times by $\gamma$, Eq.~\eqref{eq:correction-fact} (dashed green line, bin $\vect{w}_2(e)$ only, i.e.\ $r \in [1.0,1.5]\,m\,\approx D$). We notice that the compensated exposure time grows steadily in time gaining a factor $3.5\times$. 
    This is a combined effect of the passenger growth and a reduction in attention and/or difficulty in adhering to physical distancing regulations. In panel (b) we scale the exposure time by the number of passengers, i.e.\ we report $\gamma T^d_{p,f}/N$ ($d=2$). This ratio, which we further scale to its value at week 18, displays a ${\approx} 100\%$ growth between week 18 and 26, to confirm that the increment of passengers  contributes only for $150\%$ of the overall exposure time growth.}
    \label{fig: events_per_week}
\end{figure*}

\section{Pedestrian tracking setup at Utrecht Central Station}\label{ref:utrecht-intro}
We benchmark our approach considering pedestrian tracking data acquired on platform 3 at Utrecht Central station, The Netherlands (cf.\ Figure~\ref{fig: dist_on_pf}). Utrecht Central, with roughly 57 million annual users, is the nation-wide busiest railway station. Since 2017, platform 3 has been equipped with 19 commercial pedestrian tracking sensors, each of which captures 3D stereo images at $f=10$ frames per second and processes them to deliver individual tracking data in a privacy-friendly manner (cf.\ sketch in Figure~\ref{fig: dist_on_pf}). The sensor view-cones are in partial overlap, which enables the sensor network to stitch together trajectory pieces acquired by the single devices. The total area covered by the set of sensors consist of the full platform width (about $3\,$m) for $120$ linear meters next to track 5, plus the area underneath escalators and staircases connecting the platform to the central hall. This yields a covered area of approximately $450\,$m$^2$. Track 5 is among Utrecht's busiest tracks and is primarily utilized by trains heading to Amsterdam Central Station and Schiphol Airport. The complex and multi-directional crowd flows on the platform are recorded with high space- and time-resolution 24/7 since March 2017 (localization precision: $O(5-10)\,$cm, similar technology to what employed in~\cite{Heuvel2019validation}). 
In normal operation conditions, the system would  capture about $100.000$ trajectories per day while, on average, only $16.000$ trajectories per day were observed in the two months after the Covid-19 outbreak (see pre- and  during- Covid-19 crowd distribution example in Figure~\ref{fig: dist_on_pf} and crowd density histograms in Figure~\ref{fig: dist_hist_tyical_day}). This unique measurement setup gives us not only the possibility of developing and testing our approach in meaningful real-life conditions, but also to compare relevant statistical observables of the pairwise distance (RDFs), before and during the Covid-19 measures.

\section{Pedestrian radial distribution functions}\label{sec:RDF}
 In theoretical physics and molecular dynamics, the radial distribution function, $\rdf(r)$ (RDF), and the radial cumulative distribution function (RCDF), $\rcdf(r)$, are established tools to characterize the distribution of pairwise distances between particles  (see e.g.~\cite{hansen1990theory}), i.e., in our case, pedestrians.

 By definition of RCDF, for a crowd with uniform spatial density $\rho$, on average, i.e.\ in the mean-field of many realizations, the number of people, $N_\rho(r)$, at a distance \emph{up to} $r$ from a generic individual  satisfies
 \begin{equation}\label{eq:intro-RCDF}
 N_\rho(r) = \rho \rcdf(r) = \rho \int_0^r \rdf(z)\,dz,
 \end{equation}
 therefore, $\rdf(r) = \partial_r \rcdf(r)$ holds. Thus, the functions $\rdf(r)$, $\rcdf(r)$ (and derived quantities) do not carry any space/time specific information, rather they relate to average properties, depending only on mutual distances. 

For instance, in unconfined space, $\rcdf(r)$ grows as the circle area, i.e.
\begin{equation}\label{eq:circle-area}
N^{unconf}_\rho(r) = \pi r^2 \rho.
\end{equation}
In our train platform, such a ${\sim} r^2$ growth ratio is possible only when  $r$ is sufficiently smaller than the platform width. Else, we expect a ${\sim}r^1$ (linear) trend, as the area growth is bound to the platform length only. This holds until platform finite size effects come into play. In Figure~\ref{fig: dist_hist} we compare the RDF and RCDF for the two density levels highlighted in Figure~\ref{fig: dist_hist_tyical_day}. We notice a depletion in the radial distribution functions at short distances when comparing with the situation pre-outbreak. As a partial anticipation of the results of this paper, in the figure we report also the RDF discounted of short-distance family-group interactions (such interactions are allowed by present regulations). As expected,  this yields a further depletion of the RDF in the region $r \lessapprox 1.5\,$m. In the figure, we additionally report the aforementioned analytic trends and the R(C)DF functions obtained through Monte Carlo numerical simulations in a rectangular domain with the same size as the platform. We specifically consider an ensemble of simulated crowds of $N$ pedestrians; individuals have a random spatial distribution satisfying a minimal mutual distance of $0.2\,$m. The figure  reports ensemble-averaged RDFs and RCDFs which thus include small-range quadratic growth, platform-width-bound linear growth and finite size effect. Measurements well-conform with  simulations.
 Short mutual distances over extended time duration are known to increase the contagion probability: RDF and RCDF can be used to evaluate the average exposure time. A pedestrian that was on the platform for a time interval $\Delta T$, was exposed, on average, for a time
\begin{equation}\label{eq:RCDF-time-rescaling}
    T^{r_c} = \Delta T \rho \rcdf(r_c),
\end{equation}
with $r_c$ being a critical distance threshold (e.g.\ $r_c$ can be the nation-wide physical distancing requirement). Similarly, the function
\begin{equation}\label{eq:RDF-time-rescaling}
    t(r) = \Delta T \rho \rdf(r),
\end{equation}
quantifies the contribution to the total average exposure time given by peer pedestrians at distance $r$.

\section{Distance-Interaction network}\label{sec:method}
In this section, we describe our scalable framework to characterize pedestrians pairwise distances, identify family-groups and distance offenders. 

Our measurements come in the form of time-stamped trajectories. As no further information is available, such as body orientation or gaze direction~\cite{gallup2012visual}  or body size/approximate age,  our identification of family-groups relies only on mutual proximity and its time-consistency. Whenever two or more pedestrians maintain a mutual short distance consistently throughout a sufficient fraction of their trajectory, they should automatically emerge  as belonging to the same family-group. Additionally, we deem implementation simplicity, while maintaining efficiency and sufficient accuracy in identifying family-groups relations, possibly in real-time, and without complex/costly data-searches.
Hence, our approach is ``additive'' (or ``incremental'') and RDF-like information is increased, on the go, in a graph data structure (at minimal memory costs), and, without computationally-costly searches in stored records, family-groups and offenders remain identified immediately. In other words, by additivity, we stress that our data structure is built online and usable after only one time-forward pass of the trajectory data.

\subsection{Graph data structure construction}

In conceptual terms, we represent the pedestrian trajectories as individual nodes of a graph $\graph$. Each node includes information specific to the trajectories, such as overall observation time, $\tau$, source and destination. These three quantities are incremental as, respectively, $\tau$ scales with the number of frames a pedestrian is observed, the source point is the initial point of a trajectory while the destination gets constantly updated with the current position until a pedestrian leaves the measurement area.  Whenever two pedestrians, say $p_1$ and $p_2$, are observed simultaneously (i.e.\ in the same frame) and their Euclidean distance, $r = d(p_1,p_2)$, is below a critical threshold $D' > D$, we memorize (properties of) this event within the weight, $\vect{w}(e)$, of the edge $e=(p_1,p_2)$, that connects the two pedestrian-nodes $p_1$, $p_2$. Specifically, the weight $\vect{w}(e)$ aims at a discrete counterpart of the RDF ($\rdf(r)$, cf.\ Eq.~\eqref{eq:intro-RCDF}) restricted to pedestrians $p_1$, $p_2$ and with support $0\leq r \leq D'$. Similarly to the RDF, also the graph $\graph$ does not hold detailed microscopic information in space/time, such as instantaneous positions.
To focus on the identification of Corona events and disentanglement of family-groups, we set $D' = 2.5\,$m, i.e.\ one meter more than the social distance required by Dutch regulations, $D=1.5\,$m$< D'$. This aims at exploring the distance dynamics in the neighborhood of the current regulations and leaves flexibility should the regulations become stricter and require additional mutual separation.

The vector weight $\vect{w}(e)$ keeps record of the number of occurrences of distance events, $r$, after a given radial quantization (binning). In the following, we consider five evenly-sized bins with sides at 
\begin{equation}\label{eq:bins-quantization}
r=\{r_0,\ldots,r_d,\ldots\}=\{0,0.5,1.0,1.5,2.0,2.5\}\,\text{m};
\end{equation}
for the sake of readability, we also indicate, the individual components of $\vect{w}(e)$, $\vect{w}_d(e)$,  as 
\begin{align*}
    \vect{w}_0(e) = \vect{w}(e)\brkival{0.0}{0.5},\\
    \vect{w}_1(e) = \vect{w}(e)\brkival{0.5}{1.0},
\end{align*}
and so on. Hence, for each time instant in which   $d(p_1,p_2) < 0.5$, the counter  $\vect{w}(e)\brkival{0}{0.5}$ is incremented of one unit. Similarly, whenever $0.5 < d(p_1,p_2) < 1.0$, the weight $\vect{w}(e)\brkival{0.5}{1.0}$ gets incremented, \textit{etc}. Note that updating the data structure requires only all the pairwise distances (smaller than $D'$) at each frame. The choice of bin size, which regulates the quality of the approximation of the RDF of $p_1$ and $p_2$, is clearly arbitrary and needs to be a trade-off between the required resolution on the RDF and memory allowance.

Consistently with Eq.~\eqref{eq:RDF-time-rescaling}, scaling the counts $\vect{w}(e)$ by the (inverse of the)  sensors sampling frequency, $f$, yields the time duration, in seconds, pedestrians $p_1$ and $p_2$ maintain a given (quantized) distance (i.e.\ $f^{-1}\vect{w}(e)\brkival{0.5}{1.0}$ is the amount of seconds $p_1$ and $p_2$ had a distance between $0.5$ and $1.0$ meters). Hence, statistical moments of $r$, weighted by $f^{-1}\vect{w}(e)$ enable to calculate the total contact time of $p_1$ and $p_2$, their average distance and fluctuations. In all cases, the statistics are restricted to $r\leq D'$ (insights on the relevant statistical properties of the graph are left to the next subsection). Operationally, we build the graph as reported in Algorithm~\ref{alg:1}. Additionally, in Figure~\ref{fig: platform_network_graph}a, we provide a visual description of the graph in the case of a subsection of our train platform, while in Figure~\ref{fig: platform_network_graph}b we show examples of typical graphs built in time windows about 10 minute-long around the train arrivals.

\begin{algorithm}[h!]
 \KwData{Trajectories dataset, possibly live-streaming}
 \KwResult{Distance-interaction graph $\graph$}
 $\graph =$   empty graph\;
 \For{ $t$ in time }{
  \textbf{add} any trajectory, $p_i$, starting at time $t$ as a node in $\graph$, store origin\;
  \textbf{update} persistence time $\tau_{p_i}$, destination of all observed trajectory-nodes $p_i$\;
  \For{$0 < i < j \leq$  \# trajectories at time $t$}{
    \If{ $d(p_i,p_j) < D'$}{  $\vect{w}(p_i,p_j)[ q(d(p_i,p_j))  ] \mathrel{+}= 1$  }
  }
 }
 \caption{Graph construction algorithm in pseudo-code. The data structure is built streaming the trajectory data once. The function $q = q(d)$ returns the distance bin to which $d$ belongs. Hence, given the quantization in Eq.~\eqref{eq:bins-quantization}, it holds $q(0<d<0.5\,m)=0$, $q(0.5\,m<d<1.0\,m)=1$, etc.\label{alg:1}} 
\end{algorithm}

\paragraph{Extensions and variations}
In settings as train platforms, not all the areas come with the same importance or criticality. The so-called ``danger zone'', the last $80\,$cm-wide buffer region on the platform that is stepped on just before boarding a train, is an example. For our use case it is imperative to have the capability of discriminating between events happening inside and outside such an area. To achieve this, we consider two separate sets of weights on each edge: $\vect{w}^{dz}(e)$ and $\vect{w}^c(e)$, which count, respectively, the time instants a given distance below $D'$ occurs when the centroid between the two pedestrians lies in the danger zone, and otherwise. According to our previous definition, $\vect{w}(e) = \vect{w}^{dz}(e) + \vect{w}^c(e)$ holds. Figure~\ref{fig: platform_network_graph}a reflects this aspect by representing  $\vect{w}^{dz}(e)$ and $\vect{w}^c(e)$ stacked (and with different color shade).

\subsection{Approximation of the short-range RDF as edge average}
The graph is a collection of RDFs functions restricted to pairs of pedestrians. As such, within the limits of the quantization, it is richer in information than the ``global'' RDF (Eq.~\eqref{eq:intro-RCDF}). The latter, in fact, can be recovered by averaging the edge weights over the entire graph, i.e.\ by combining each pairwise contribution. Restricting to a graph describing conditions with equal density, the global RDF can be approximated as
\begin{equation}\label{eq:short-range-RDF-approx}
    \int_{r_d}^{r_{d+1}}\rdf(r)\, dr \approx c\bigg\langle \frac{\vect{w}_d(e)}{\sum_j \vect{w}_j(e) } \bigg\rangle_{e},\ \forall d,
\end{equation}
where $c$ is a constant scaling depending on the normalization considered for $g$.
In words, the integral of the RDF in the bin $[r_d,r_{d+1}]$ can be approximated by the $d$-th (enseble-averaged) edge weight. Averaging over a graph including non-homogeneous density levels, yields the RDF averaged among such densities.  

\subsection{Interaction classification}
In this subsection we leverage on the graph topology and edge data to deduce relevant information about pairwise distance, family-group relations, exposure times, and physical distance offenders.

\paragraph{Pairwise exposure time and pairwise distance statistics}
Pedestrians $p_1$ and $p_2$, whose distance satisfied $r = d(p_1,p_2) \leq D'$ at least in one frame, have their interaction recorded on the graph edge $e=(p_1,p_2)$. The weight $\vect{w}(e)$ allows us to characterize their distance properties restricted to the instants in which $r<D'$. In particular, the \textit{contact time} $T_e$ between $p_1$ and $p_2$ satisfies
\begin{equation}\label{eq:pairwise-contact-time}
    T_e^d = f^{-1}\sum_{j = 0}^{d} \vect{w}_j(e), \qquad \mbox{[Contact time]}
\end{equation}
where the index $d$ selects the farthest relevant distance bin. According to our quantization in Eq.~\eqref{eq:bins-quantization}, $d=2$ would restrict to interactions with $r\leq 1.5\,$m and thus quantify the ``exposure time'' according to the Dutch regulations, whereas $d = d_{max}=4$ includes all stored interactions for the pair $e$, i.e.\ $r\leq D'$.
Similarly, the average pairwise distance reads
\begin{equation}
    \langle r \rangle_{e}^d = \frac{f^{-1}\sum_{j = 0}^{d} r_{j + \frac{1}{2}} \vect{w}_j(e)}{T_e^d}, \qquad \mbox{[Avg. pairwise dist.]}
\end{equation}
where $r_{j + \frac{1}{2}}$ identifies the mid point between bin $j$ and $j+1$ in Eq.~\eqref{eq:bins-quantization}. Higher order moments of $r$ weighted by $\vect{w}(e)$ can be used to estimate the fluctuation (variance) of the distance in time. In particular, if the variance $\sigma^2(r) = \langle r^2 \rangle_e^d - (\langle r \rangle_e^d)^2$ is small, then the pedestrians kept an almost fixed distance during their trajectories. Notably, small or possibly zero $r$ variance, $\sigma^2(r)$, are necessary consition for non-positive (Finite Time) Lyapunov Exponent of the distance between $p_1$ and $p_2$~\cite{vallejo2017predictability}. %

\paragraph{Total individual exposure time}
The total time an individual $p$ has been exposed to contacts can be computed by summing the pairwise contact times $T_e^d$ (Eq.~\eqref{eq:pairwise-contact-time}) for all pedestrians, $p_j$, that entered into contact with $p$, i.e.\ for all the edges $e=(p,p_j)$ that converge to $p$, in formulas
\begin{equation}\label{eq:total-no-family-contact-time}
    T_p^d = \sum_{p_j \in N(p)} T_{(p,p_j)}^d,  \qquad \mbox{[Individ. expos. time]}
\end{equation}
where $N(p)$ is the list of the first-neighbor of $p$ (nodes connected to $p$ through at least a single edge). Equation~\eqref{eq:total-no-family-contact-time} provides a counterpart to Eq.~\eqref{eq:RCDF-time-rescaling} in which we consider a specific pedestrian, $p$, rather than averaging over all pedestrians. Notice that the index $d$ is the discrete analogue of the cutting threshold $r_c$.

\paragraph{Family-group relations} 
We determine whether two pedestrians, $p_1$, $p_2$, belong to the same family-group on the basis of their contact time $T^d_{(p_1,p_2)}$ and their persistence time in the tracking area, $\tau_{p_1}$ and $\tau_{p_2}$. In particular, if the symmetric relation henceforth indicated as $p_1\sim p_2$ holds 
\begin{align}\label{eq:family-rel}
&& p_1\sim p_2 \iff  \quad\qquad \mbox{[Fam-group condition]}\\
&&\text{min}\left(\frac{T^{(1)}_e}{\tau_{p_1}},\frac{T^{(1)}_e}{\tau_{p_2}}\right) > \lambda^{(1)}  \quad \text{and} \quad \text{min}\left(\frac{T^{(2)}_e}{\tau_{p_1}},\frac{T^{(2)}_e}{\tau_{p_2}}\right)  > \lambda^{(2)},\nonumber %
\end{align}
we consider $p_1$ and $p_2$ as belonging to the same family-group. We set $\lambda^{(1)} = 40\%$, $\lambda^{(2)}  =90\%$ which, in words, translates to
people who have a pairwise distance of less than $1.5\,$m for $90\%$ percent of the time and are within $1\,$m for $40\%$ percent of the time. The rationale being that pedestrians who followed the same trajectory, thereby being in mutual close proximity for the major part of their persistence time, and who are comfortable for extended periods in each other's private space ($r\leq1\,$m) most likely belong to the same family-group. Family-groups with more than two individuals are expected to appear in the graph as completely connected sub-graphs, or cliques, in which all the nodes are in Relation~\eqref{eq:family-rel} between each other. 

We can now define a second total individual exposure time which is equal to $T_p^d$ (Eq.~\eqref{eq:total-contact-time}) after discounting family-group contacts:
\begin{equation}\label{eq:total-contact-time}
    T_{p,f}^d = \sum_{\substack{p_j \in N(p) \\ p_j \nsim p}}  T_{(p,p_j)}^d.  \qquad \mbox{[Individ. expos. time w/o families]}
\end{equation}
Analogously, we can consider a RCDF discounted of family-group contributions, say $\rcdf_f(r)$ (cf.\ Eqs.~\eqref{eq:intro-RCDF},~\eqref{eq:RCDF-time-rescaling}), such that 
\begin{equation}
    T^{r_c}_f = \Delta T \rho \rcdf_f(r_c)
\end{equation}
is the total exposure time with non-family individuals and up to a spatial threshold $r_c$. By differentiation (as in Eq.~\eqref{eq:intro-RCDF}), we can similarly define the RDF $g_f(r)$ discounted by family-groups. 

\paragraph{Family sub-graph transitive closure}
The relation ``$\sim$'' can be non-transitive, i.e.\  $p_1{\sim}p_2$, $p_2{\sim} p_3$ do not imply $p_1{\sim}p_3$. One may thus consider the transitive closure of ``${\sim}$'', say ``$\bar{\sim}''$ which is defined as $p_1$ and $p_3$ belong to the same family-group ($p_1{\bar{\sim}} p_3$) either if $p_1\sim p_3$ or if they have common family-group members. 

In the data we analyze in  Section~\ref{sect:results}, we do not consider such transitive closure. In other words, family-group relations are exclusively defined by Relation~\eqref{eq:family-rel}.  On one hand, we deem rare the event that a family remains not represented by a clique. Even in this case, we expect contributions to the overall RDF statistics be minimal. On the other hand, in real-life data collected with sensors similar to ours, short/broken trajectories may appear. We observed that these, in combination with Relation~\eqref{eq:family-rel} would unrealistically increase the probability of observing large family-groups. Hence, to avoid excessive and unjustified complications in the heuristics we restrict to Relation~\eqref{eq:family-rel}.

\begin{figure*}[t]
    \centering
    a. \includegraphics[width=0.43\linewidth]{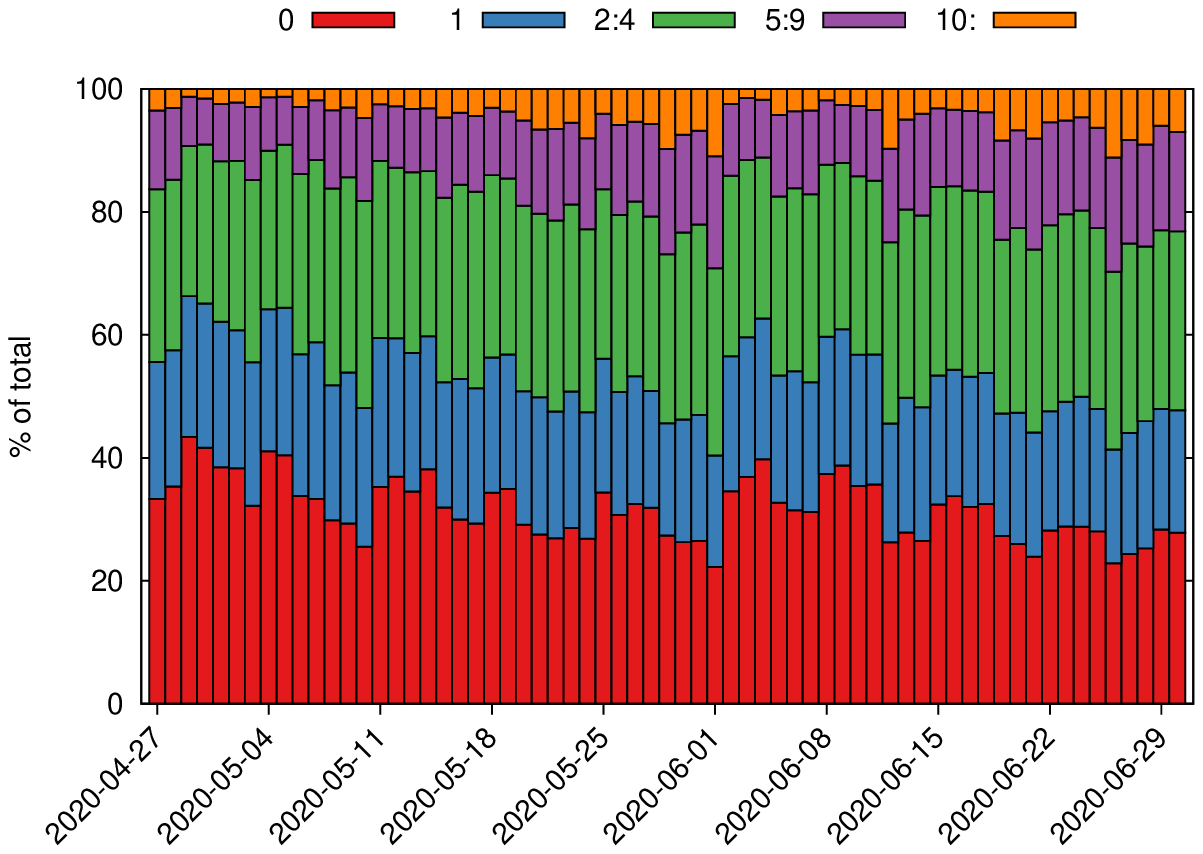} %
    b. \includegraphics[width=0.45\linewidth]{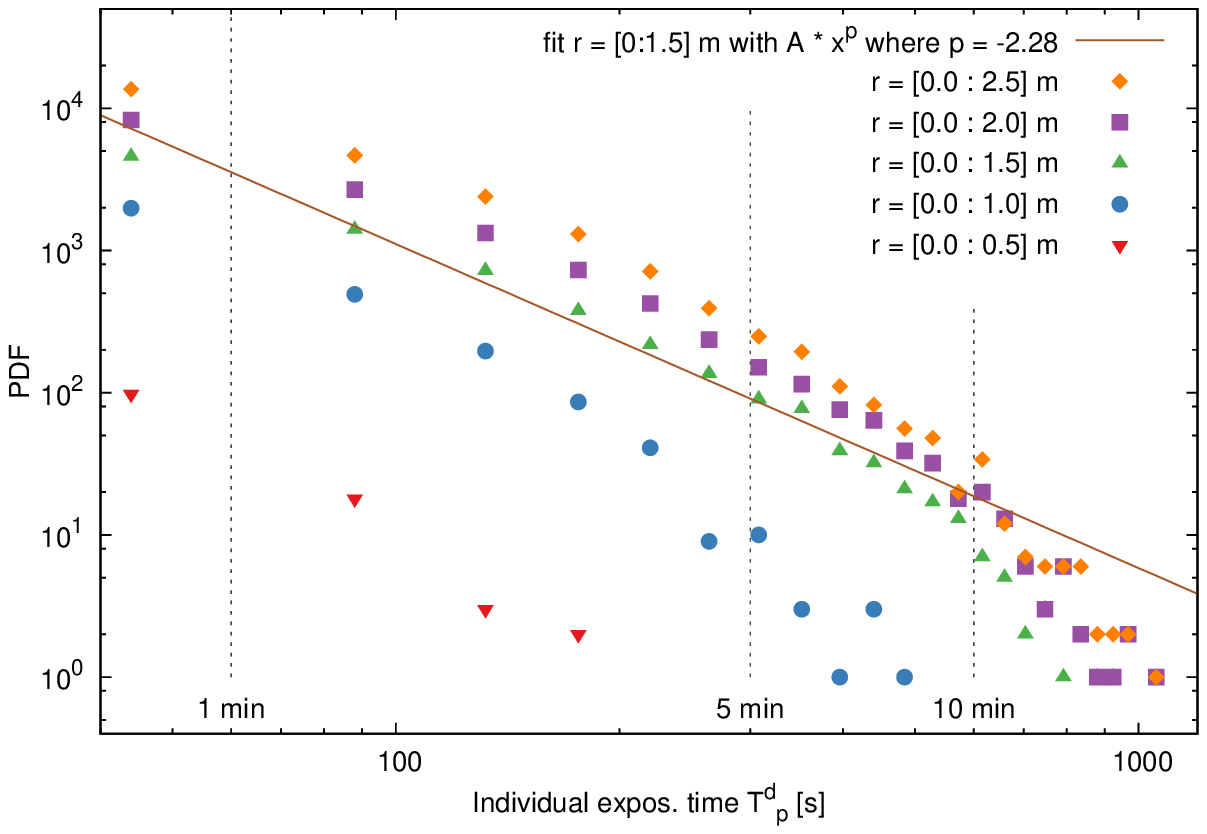}    %
    \caption{(a) Distribution of node-pedestrian degree per day as a percentage of the total number of passengers. The degree of a node counts the number of people encountered with a mutual distance smaller than $1.5\,$m (hence, degree $0$ means that a person did not have any Corona event). We observe that high-degree nodes, i.e.\ repeated distance offenders, increased steadily until the train schedule change (e.g.\ nodes with 10+ contacts grew from ${\approx} 1\%$ to ${\approx} 10\%$). The schedule change yielded a temporary drop in the offender percentage after which it started increasing again. This can be a sign of warning towards the relaxation in the compliance of physical distancing rules.
    (b)  Probability density function of the individual exposure time discounted of families, $T^d_{p,f}$ considering different maximum distances (Eq.~\eqref{eq:total-no-family-contact-time}). Exposure times show a power-law behavior. The PDF depletion after $T = 5\,$minutes is most likely due to the time windowing that we operate around each train arrival (cf.\ Figure~\ref{fig: platform_network_graph}b). This yields a data cut-off for large times.
    }
    \label{fig: dist_hist_discounted_month}
\end{figure*}

\begin{figure*}[t]
    \centering
    a.\includegraphics[width=0.47\linewidth]{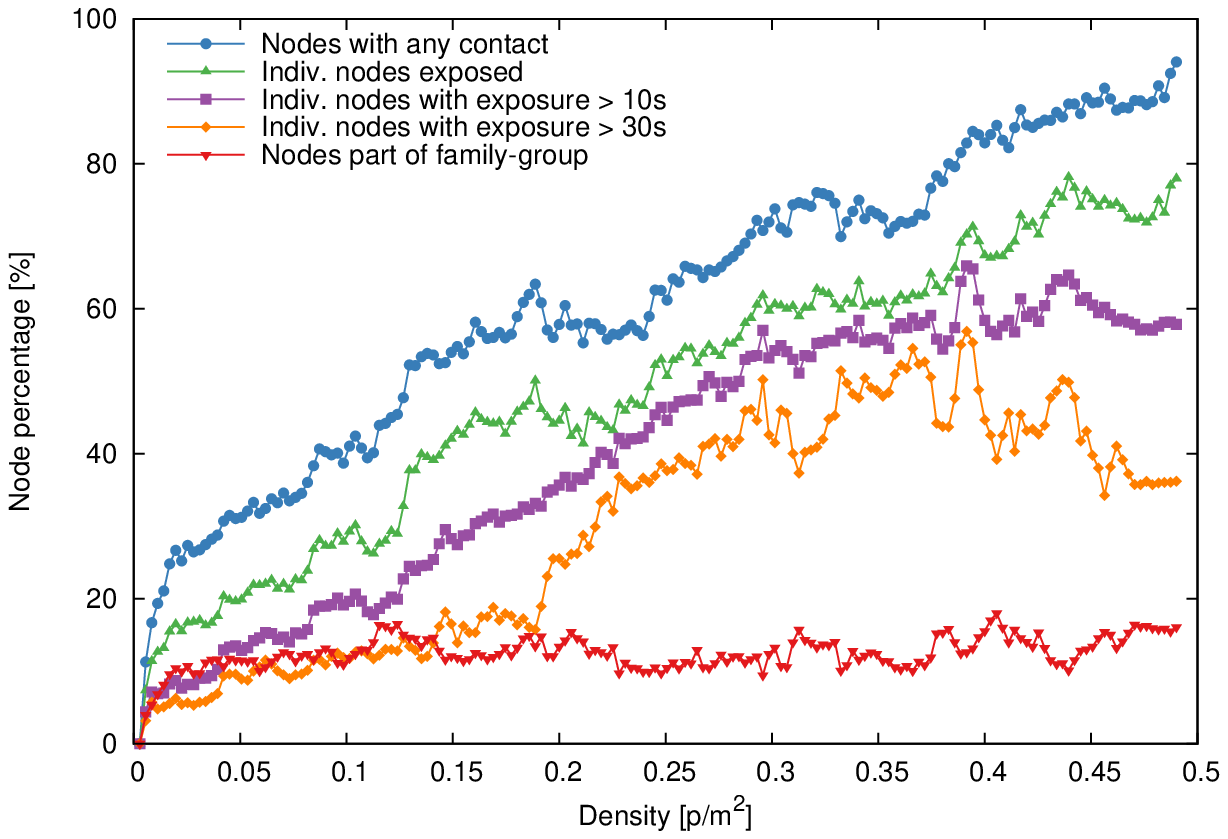}
    b. \includegraphics[width=0.4\linewidth]{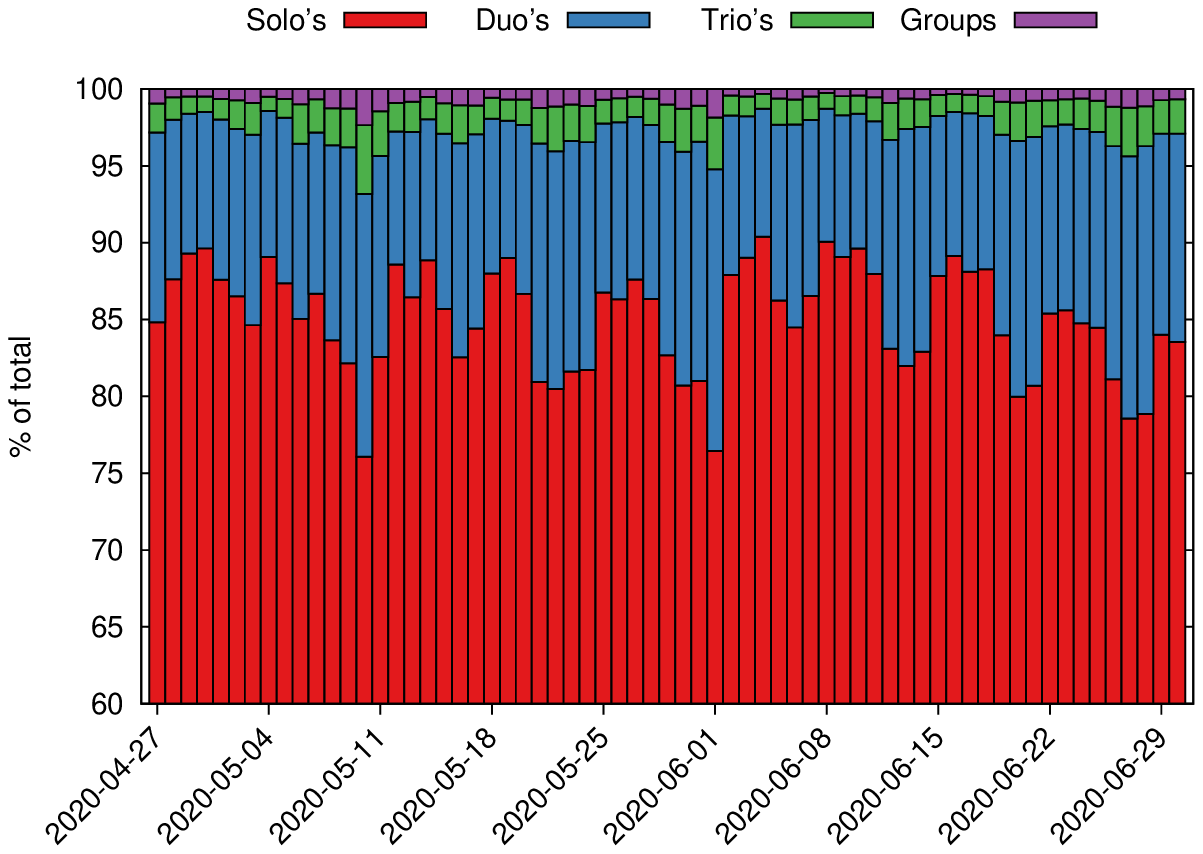}
    \caption{(a) Percentage of pedestrian nodes exposed to contacts as a function of the global density on the platform (density calculated as number of people in a frame divided by the total sensor area, $450\,$m$^2$, discounted of the danger zone, $96\,$m$^2$). Exposed nodes that have at least one contact, of any duration, with another pedestrian (within or outside their family-group or not) are in blue. This percentage if further broken down into nodes part of a family group (red) and actual distance offenders (green). The purple and orange lines restrict, respectively, to nodes with a minimum exposure time of $10\,$s and $30\,$s. Linear fitting parameters are reported in Table~\ref{tab:fitting}. (b) Distribution of individuals and cliques day-by-day as a percentage of the total number of nodes. Between $80\%$ and $85\%$ of the nodes do not belong to cliques, i.e.\ they travel alone and their contacts are all distance infringements. Family-groups of two people cover about $12\%-15\%$ of the remaining nodes; family-groups of three and more provide a minimal ${\approx}3\%$ contribution.   %
    }
    \label{fig: dens_vs_offenders}
\end{figure*}

\paragraph{Relevant interactions, family-discounted graph and offenders}
Combining the previous elements we can now identify distance offenders as those pedestrians that violate physical distancing while not being part of a family. We consider the sub-graph $\graph' \subset \graph$ obtained after pruning $\graph$ of family-group edges. The connections left in $\graph'$ must indicate sporadic (i.e.\ not time-consistent) distance infringements.  

As exposure time is deemed a critical parameter for contagion~\cite{WHO}, we apply a further time requirement to discriminate actual offenders. Specifically, we introduce the set $P'_{\alpha}$ defined as
\begin{equation}\label{eq:Pprime_set}
    P'_{\alpha} = \{ p \in \graph' \colon T^d_{p,f} > \alpha \};
\end{equation}
in words, elements of $P'_{\alpha}$ are pedestrians who violated physical distancing with non-family members for an overall time longer than $\alpha$. The number of first neighbors of a node in $P'_{\alpha}$ identifies how many contacts such pedestrian had: we label as repeated offenders those with more than 10 first neighbors (i.e.\ pedestrians that violated physical distancing with more than 10 different people and for an overall time larger than $\alpha$). We remark that this classification can be run in real-time as all the aforementioned requirements can be constructed in additive manner.

\section{Physical distancing at Utrecht Central Station,\\ platform 3}\label{sect:results}
In this section we employ the graph $\graph$ to analyze trajectory data collected in Utrecht station (see Section~\ref{ref:utrecht-intro}) and we compare statistics from before and during the Covid-19 pandemic.

\paragraph{Typical graphs and qualitative aspects}
In Figure~\ref{fig: platform_network_graph}b, we report examples of the graphs acquired during a typical morning (time interval 4AM - 8.30AM). Train arrivals are the most critical conditions when it comes to respecting physical distancing, thus we create a new graph two minutes after each train departure, when the platform is almost empty (this step is not strictly necessary, but increases computational efficiency). In the figure, nodes size and color follows the node degree, i.e.\ the number of first neighbors and, thus, the distance offenses committed by that node. 

As a qualitative example of the capability of the method to extract relevant data, we showcase two antipodal conditions in Figure~\ref{fig: duo_offender}. In the first case (left panel), we report two pedestrians in a family-group relation that remain together throughout their entire trajectories: from the escalators to the boarding. In the second case, we have a repeated offender: the associated node exhibits $28$ first-neighbor connections. Interestingly, a significant part of the offenses happens while the pedestrian waits in proximity of the escalator. This, therefore, rather marks a waiting area to be disallowed, than a willing offender.

\paragraph{Family-group discounted RDFs}
In Figures~\ref{fig: dist_hist} and~\ref{fig: dist_hist_weekend}  we report RDFs prior and after excluding family-group interactions. The RDFs for $r<D$ are non-vanishing, even after discounting the contributions of family-groups, most significant in the weekends  (when the presence of workers and commuters is lower; cf.\ short-distance ``bump'' in Figure~\ref{fig: dist_hist_weekend}). We expect these remaining contributions at $r<D$ to be to Corona events by distance offenders. 
Notably, once removed of family-groups contributions, the RDFs at small $r$ values recover a linear growth rate, as expected by a random spatial distribution of passengers (scaling as the derivative of Eq.~\eqref{eq:circle-area}).

\paragraph{Exposure time, node degree and evolution through the pandemics}
In Figure~\ref{fig: events_per_week}a we report, week-by-week, the average edge weight as introduced in Eq.~\eqref{eq:short-range-RDF-approx}, pruned of family-group contributions and scaled by the sampling frequency. This provides an approximation of how the (family-discounted) average individual exposure time has been changing over time (weeks 17 to 26 in 2020). As the usage of the platform grew after a drop at the beginning of the outbreak, so did the exposure time for distances between $0.5\,$m and $2.5\,$m, especially until week 22. On the opposite, the amount of time, per person, spent with a peer within $0.5\,$m has remained constant and within fractions of a second.  A new operation schedule at week 23, with increased train frequency throughout the day, allowed a temporary reduction of the load on the platform, making easier to respect physical distancing. From week 24 onward, individual exposure times showed again a growing trend. To render the data comparable, we consider also exposure times compensated for the new train schedule (Figure~\ref{fig: events_per_week}a, only for the case $r\approx D$) , i.e.\ corrected by a factor 
\begin{equation}\label{eq:correction-fact}
\gamma = \frac{\#\,\text{daily trains with new schedule}}{\#\,\text{daily trains with old schedule}}\approx 1.7.
\end{equation}
This shows a more stable growth pattern and an increase of factor $3.5$ between weeks 17 and 26 (the factor $\gamma$ is an estimate, considering the presence of trains of different kind and lengths). Scaling the corrected exposure times with the number of passengers, which is itself growing, we additionally notice, that the former is growing faster (i.e. exposure time grow super-linearly with respect to the passengers). This suggests a possible relaxation or an increased difficulty in following physical distancing regulations.

We report in Figure~\ref{fig: dist_hist_discounted_month}a an in-depth breakdown of the distribution of node degrees, i.e.\ the number of first neighbors of each node and thus the number of contacts with different individuals the node had (including both offenses and families). Consistently with our previous observations, the fraction of high degree nodes (5+ or 10+), i.e.\ repeated offenders, has also been growing steadily, but a temporary drop following the train schedule change. 
In Figure~\ref{fig: dist_hist_discounted_month}b we display the distribution of individual exposure times pruned of family contributions, and up to the distance thresholds $r_d$ (Eq.~\eqref{eq:bins-quantization}), i.e.\ the pdfs of $T^d_{p,f}$ (cf.\ Eq.~\eqref{eq:total-contact-time}). Similarly to what discussed in~\cite{cattuto2010dynamics}, and consistently with the model in~\cite{karagiannis2010power}, we observe a power-law distribution in the exposure times (exponent $p < -2$), which emerges as a robust feature of random encounter dynamics. Additionally, we notice that the distance threshold plays a strong multiplicative effect and, possibly, it also weakly influences the exponent. It is worth remarking that our largest observation times are bound by the fact that we limit our graphs to time intervals of about 10-15 minutes around each train arrival. This reduces our resolution at large time scales and thus yields the exponential-like drop in the distribution tails.

\paragraph{Crowd density: incidence of family-groups and effective offenders}
The passenger density on the platform has an influence on the offenses: the higher the density the easier it gets to violate physical distancing. In Figure~\ref{fig: dens_vs_offenders}a we report for a sample day (12$^{th}$ of June 2020), how the percentage of ``family'' nodes and offenders scales with the density within the global density interval  $[0,0.5]\,$ped/m$^2$.
As in Eq.~\eqref{eq:Pprime_set}, we also include minimum contact time thresholds, $\alpha$, for tagging offenders. We observe that the percentage of nodes belonging to family-groups remains stationary (value ${\approx}11\%$, a detailed breakdown of the clique size is in Figure~\ref{fig: dens_vs_offenders}b). Up to $80\%$ of the pedestrians at the maximum density level committed offenses assuming no minimum time threshold ($\alpha = 0$). This percentage slightly diminishes to $60\%$ restricting to a minimum contact time $\alpha = 5\,$s. Interestingly, the percentage shows a non-linear dependency on the density when $\alpha = 30\,$s. In particular, such percentage remains stationary  (value $\approx 11\%$) until $0.2\,$\pedsq\ and then suddenly increases. This can suggest an increase in difficulty in following distancing rules around this density level. We report the coefficients of the linear fitting of such data in Table~\ref{tab:fitting}.

\begin{table}[t]
\caption{Linear fit parameters for the node classification (percentage data) in Figure~\ref{fig: dens_vs_offenders}a. As a  measurement unit for the pedestrian density we employ \textit{tenths of pedestrians per square meter}: $\delta = 10^{-1}\text{ped$/$m}^{2}$. Thus the fitting intercept is at $\delta = 0$ while the increments (slopes) are reported as percent variations per $\delta$ unit. As an example, pedestrians that do not belong to a family-group and are in contact with someone else grow about $13\%$ when the density increases between $0.2$ and $0.3\,$ped/m$^2$ and similarly between $0.3$ and $0.4$ and so on. We do not report linear fitting parameters for the case $\alpha = 30\,$s as the growth is non-linear.
\label{tab:fitting}}
\begin{tabular}{l | l | l}
                                       &  $\delta \approx 0$ & Slope \\ \hline
Ped.-nodes with any contact                        &   28.5$\,\%$      & 13.4$\,\%/\delta$   \\
Ped.-nodes exposed                 &  17.1$\,\%$       & 12.9$\,\%/\delta$   \\
Ped.-nodes with exposure $\alpha > 10\,$s & 9.56$\,\%$ & 12.4$\,\%/\delta$   \\
Ped.-nodes part of a family-group                  & 11.5$\,\%$   & 0.48$\,\%/\delta$ 
\end{tabular}
\end{table}

\section{Discussion}\label{sect:discussion}
We have presented an highly efficient and accurate approach to the problem of studying, in real-time, the distance-time encounter patterns in a crowd of individuals. Our approach allows us to identify social groups, such as families, by imposing thresholds on the distance-time contact patterns. In the context of the currently ongoing Covid-19 pandemic, we demonstrate this as an effective and promising tool to monitor, in a full privacy respectful way, the observation of physical distancing. The outcome of the analysis can provide early warnings in respect to an average relaxation towards the compliance of physical distance rules, can allow to identify spots where physical distancing most frequently is violated and it may, as well, allow to identify in real-time the presence of distance offenders. We observed, besides, a super-linear dependence between contact times and passenger number. This can be caused both by a reduction of attention towards social distancing rules but also to an intrinsic increase in difficulty in complying to regulations. The investigation of this aspect is left to future research.

The proposed algorithm is simple and can be easily implemented using existent graph code libraries. In our case, we could process a day of data in few minutes using the python NetworkX library~\cite{hagberg2008exploring}. Libraries sporting higher performance and/or scalability exist in case of even more demanding situations.

It is worth mentioning that the approach here proposed can be applied to any type of tracing trajectories and possibly to study the collective dynamics of large groups of active or passive particles making it a tool capable of going well beyond the application to crowd dynamics and physical distancing discussed here.

\bibliography{references}

\end{document}